\documentclass[12pt]{iopart}
\usepackage{iopams}
\eqnobysec

\begin{document}
\jl{1}
\newtheorem{thm}{Theorem}

\title[A list of all integrable 2D homogeneous polynomial potentials]
{A list of all integrable 2D homogeneous polynomial potentials with 
a polynomial integral of order at most 4 in the momenta
}

\author{Katsuya Nakagawa\dag\footnote[3]{e-mail: k.nakagawa@nao.ac.jp} 
and Haruo Yoshida\ddag\footnote[4]{e-mail: h.yoshida@nao.ac.jp}}

\address{\dag\ Department of Astronomical Science, 
The Graduate University for Advanced Studies, 
Mitaka, Tokyo 181-8588, Japan}

\address{\ddag\ National Astronomical Observatory of Japan, 
Mitaka, Tokyo 181-8588, Japan}

\begin{abstract}
We searched integrable 2D homogeneous polynomial potential 
with a polynomial first integral by using the so-called direct 
method of searching for first integrals. 
We proved that there exist no polynomial first integrals 
which are genuinely cubic or quartic in the momenta if the 
degree of homogeneous polynomial potentials is greater than 4. 
\end{abstract}

\section{Introduction}\label{sec:intro}
\noindent
A Hamiltonian system with \(n\) degrees of freedom is integrable 
if the system admits \(n\) independent first integrals in involution 
(Liouville integrability). 
It is a fundamental and important problem to investigate the 
integrability of Hamiltonian systems. 
Since the Hamiltonian itself is a first integral, the case of 
one degree of freedom is trivial. 
The simplest but non-trivial problem arises in the case of 
two degrees of freedom. 
For this case, the existence of an additional first integral 
guarantees the integrability. 
Let us consider a Hamiltonian system with two degrees of freedom, 
\begin{equation}\label{eqn:system}
H = \frac12 (p_x^2 + p_y^2) + V(x,y).
\end{equation}
At present, there is no ultimate algorithm 
(necessary and sufficient conditions for integrability) 
to determine whether the system (\ref{eqn:system}) is 
integrable or not for a given potential \(V(x,y)\). 

The solution of the system is said to possess the Painlev\'e 
property if it has no movable singular point other than poles. 
The Painlev\'e property of the solution has been believed to be 
closely related to the integrability of the system, which is known 
as the Painlev\'e conjecture (see, for example, 
\cite{kruskal1990,kruskal1992,ramani1989}). 
Although any rigorous relation between the Painlev\'e property and 
the integrability has not been established, some new integrable 
systems were detected~\cite{gramma1982,ramani1982} by postulating 
that the solution possesses the Painlev\'e property (the Painlev\'e test). 

For some Hamiltonian systems of the form (\ref{eqn:system}), 
it is possible to prove the non-integrability, i.e. the non-existence 
of an additional first integral. 
In the early 1980s, Ziglin~\cite{ziglin1983a,ziglin1983b} presented 
a non-integrability theorem and proved the non-integrability of some 
well-known Hamiltonian systems. 
Yoshida~\cite{yoshida1987} gave a criterion for non-integrability of 
Hamiltonian systems (\ref{eqn:system}) with a homogeneous 
potential by using Ziglin's theorem~\cite{ziglin1983a}. 
Recently, the differential Galois theory has become an important tool 
in attacking the problem of integrability 
(see, for example, \cite{baider1996,church1995}). 
Morales-Ruiz and Ramis~\cite{morales2001c} obtained a stronger 
necessary condition for integrability from their own 
theorem~\cite{morales2001a,morales2001b} based on the 
differential Galois theory (see also \cite{morales1999}). 
This necessary condition also justified the so-called weak-Painlev\'e 
property~\cite{ramani1982,ramani1989} as a necessary condition for 
integrability for the first time~\cite{yoshida1999}. 

On the other hand, we have difficulty in proving the integrability. 
Even though a given system passes the Painlev\'e test or satisfies 
the necessary condition for integrability, it remains unknown whether 
or not the system is actually integrable. 
In order to prove the integrability of the system, we have to present 
a first integral which is independent of the Hamiltonian. 
However, there is no general method to obtain an explicit expression 
of the desired first integral. 
Let \({\mathit \Phi}\) be a first integral of the system (\ref{eqn:system}). 
Then the Poisson bracket of \({\mathit \Phi}\) and \(H\) vanishes, 
which gives the following partial differential equation (PDE). 
\begin{equation}\label{eqn:poisson}
\fl \frac{\partial {\mathit \Phi}}{\partial x} 
\frac{\partial H}{\partial p_x} -
\frac{\partial {\mathit \Phi}}{\partial p_x} 
\frac{\partial H}{\partial x} +
\frac{\partial {\mathit \Phi}}{\partial y} 
\frac{\partial H}{\partial p_y} -
\frac{\partial {\mathit \Phi}}{\partial p_y} 
\frac{\partial H}{\partial y} = 
p_x \frac{\partial {\mathit \Phi}}{\partial x} -
\frac{\partial {\mathit \Phi}}{\partial p_x} 
\frac{\partial V}{\partial x} +
p_y \frac{\partial {\mathit \Phi}}{\partial y} -
\frac{\partial {\mathit \Phi}}{\partial p_y} 
\frac{\partial V}{\partial y} = 0.
\end{equation}
In the present paper, we search polynomial solutions of the PDE 
(\ref{eqn:poisson}), i.e. we assume that the first integral is a polynomial 
in \((x,y,p_x,p_y\)) and that the potential is a polynomial in \((x,y)\). 
The advantage of considering polynomials is that the PDE 
(\ref{eqn:poisson}) becomes an identity for \((x,y,p_x,p_y)\) and that
the problem reduces to completely algebraic one. 
In addition, we assume that the potential \(V(x,y)\) is a 
homogeneous polynomial, motivated by the following three points: 
(i)~Any polynomial potential \(V(x,y)\) can be written in the form of 
the sum of homogeneous parts as
\begin{equation}\label{eqn:polynomial}
V(x,y) = V_{\rm min} (x,y) + \cdots + V_{\rm max} (x,y),
\end{equation}
where \(V_{\rm min}\) and \(V_{\rm max}\) are the lowest degree 
part and the highest degree part, respectively; 
and it can be shown~\cite{hieta1987} that if the system 
(\ref{eqn:system}) with the potential (\ref{eqn:polynomial}) admits 
a polynomial first integral, then the systems only with the lowest 
degree part and the highest degree part, given by
\begin{equation}\label{eqn:homogeneous}
H = \frac{1}{2} (p_x^2 + p_y^2) + V_{\rm min} (x,y), \quad
H = \frac{1}{2} (p_x^2 + p_y^2) + V_{\rm max} (x,y)
\end{equation}
both admit polynomial first integrals. 
Namely, in order for the system with a non-homogeneous 
potential to be integrable, each of the systems only with the highest 
degree part of the potential and the lowest degree part of the 
potential must be integrable. 
(ii)~As we will see in section~\ref{sec:state}, 
the homogeneity of potentials assumes the weighted-homogeneity 
of first integrals, which simplifies the form of first integrals and 
reduces the complexities of computations. 
Indeed, the computations for non-homogeneous potentials are too 
complicated to deal with. 
(iii)~The homogeneity of potentials plays an essential role 
to obtain the non-integrability criterions in 
\cite{morales2001c,yoshida1987,yoshida1999}. 
For these reasons, we treat Hamiltonian systems (\ref{eqn:system}) 
with a homogeneous polynomial potential of degree \(k\),
\begin{equation}\label{eqn:potential}
V(x,y) = V_k = \sum_{j=0}^k \alpha_j x^{k-j} y^j 
= \alpha_0 x^k + \alpha_1 x^{k-1} y + \cdots + \alpha_k y^k.
\end{equation}
Let us here mention the rotational degrees of freedom. 
The integrability is preserved under rotations of coordinates, 
i.e. a potential obtained from an integrable potential by a rotation 
of coordinates is again integrable. 
We should identify such two potentials. 
In the present paper, we assume that \(\alpha_1=0\) from the 
beginning, which partially removes the rotational degrees of freedom. 

Hietarinta~\cite{hieta1983} performed the direct search for 
integrable systems of the form (\ref{eqn:system}), 
where the potential is a homogeneous polynomial of degree 5 or less 
and the additional first integral is a polynomial of order 4 or less in 
the momenta. 
The purpose of the present paper is to extend the degree of 
homogeneous polynomial potentials to an arbitrary positive integer 
in order to make a complete list of integrable systems in the range 
studied. 

We have to admit that our search does not cover the whole of 
integrable systems. 
In fact, there are some integrable systems with a rational or 
transcendental first integral~\cite{hieta1984,hieta1987}. 
More generally, as seen in the approaches based on Ziglin's analysis 
or the differential Galois theory, integrability requires meromorphic 
first integrals.
However, it would be almost impossible to single out all 
integrable cases without any restrictions on the class of the 
first integral and of the potential. 
It may be said that restrictive assumptions are 
indispensable for carrying out a thorough search for integrable 
systems. 
In this sense, our setting, in which the whole computations are 
tractable, is the fundamental one for a thorough search. 

This paper is organized as follows. 
The new result (theorem~\ref{thm:main}) will be shown after a brief 
summary of the known facts in section~\ref{sec:known}. 
In section~\ref{sec:state}, the statement of theorem~\ref{thm:main} 
is amplified, followed by details of the computations 
in section~\ref{sec:detail}. 
Finally, in section~\ref{sec:list}, we present a list of all integrable 
2D homogeneous polynomial potentials with 
a polynomial first integral of order at most 4 in the momenta.

\section{The known facts and the new result}\label{sec:known}
\noindent
The classification of first integrals up to quadratic in the momenta 
is well-known~\cite{hieta1987}. 
The existence of a first integral linear in the momenta is related with 
symmetry, the invariance of the system under a transformation: 
the conservation laws of momentum and angular momentum correspond to 
the invariances by rotations and translations, respectively. These are 
particular cases of Noether's theorem~\cite{arnold1989,hill1951}. 
First integrals quadratic in the momenta are exhausted by 
the Bertrand-Darboux theorem: 
\begin{thm}[\cite{marshall1988}]
The following three conditions are equivalent. \\
(i) A Hamiltonian system of the form $(\ref{eqn:system})$ possesses 
an additional first integral quadratic in the momenta. \\
(ii) The potential function satisfies a linear partial differential 
equation called Darboux equation. \\
(iii) The system is separable in elliptic, polar, parabolic, or 
Cartesian coordinates.
\end{thm}
If we apply the Bertrand-Darboux theorem to the potential 
(\ref{eqn:potential}), then we obtain the following homogeneous 
polynomial potentials with an additional polynomial first integral 
quadratic (or linear) in the momenta.
\begin{itemize}
\item separable in polar coordinates
\numparts
\begin{eqnarray}
V_k = r^k = (x^2 + y^2)^{k/2} 
= \sum_{m=0}^{k/2} {k/2 \choose m} x^{k-2m} y^{2m}, 
\quad k=\mbox{even} 
\label{eqn:polar} \\
{\mathit \Phi} = y p_x - x p_y
\label{eqn:ipolar}
\end{eqnarray}
\endnumparts
\item separable in parabolic coordinates~\cite{ramani1982}
\numparts
\begin{eqnarray}
\fl V_k = \frac{1}{r} \left[ \left (\frac{r+x}{2} \right)^{k+1} 
+ (-1)^k \left (\frac{r-x}{2} \right)^{k+1} \right]
= \sum_{m=0}^{[k/2]}2^{-2m} {k-m \choose m} x^{k-2m} y^{2m}
\label{eqn:parabolic} \\
\fl {\mathit \Phi} = p_y (y p_x - x p_y) + \frac{1}{2} y^2 V_{k-1}
\label{eqn:iparabolic}
\end{eqnarray}
\endnumparts
\item separable in Cartesian coordinates
\numparts
\begin{eqnarray}
V_k = A x^k + B y^k
\label{eqn:Cartesian} \\
{\mathit \Phi} = p_x^2 + 2 A x^k \quad 
(\mbox{including \({\mathit \Phi} = p_x\) for \(V_k = y^k\)})
\label{eqn:iCartesian}
\end{eqnarray}
\endnumparts
\end{itemize}
Note that the potential separable in elliptic coordinates drops out 
because it can not be a homogeneous polynomial. 
It is also possible to obtain the above list by using the direct search 
(see \cite{dorizzi1983,hieta1987}). 

There seems to be no special property of the system connected to 
the existence of first integrals of higher orders in the momenta. 
We can make polynomial first integrals of {\em apparently} 
higher orders in the momenta from the above polynomial first 
integrals (\ref{eqn:ipolar}), (\ref{eqn:iparabolic}), and 
(\ref{eqn:iCartesian}). For example, 
\begin{equation}\fl
{\mathit \Phi} = (y p_x - x p_y)^4, \quad
{\mathit \Phi} = \left\{p_y (y p_x - x p_y) 
+ \frac{1}{2} y^2 V_{k-1}\right\}^2, \quad
{\mathit \Phi} = (p_x^2 + 2A x^k)^2
\end{equation}
are polynomial first integrals which are {\em apparently} quartic 
in the momenta for the potentials (\ref{eqn:polar}), 
(\ref{eqn:parabolic}), (\ref{eqn:Cartesian}), respectively. 
On the other hand, 
Hall~\cite{hall1983} and Grammaticos~\etal~\cite{gramma1982} 
found independently the potential of degree 3,
\numparts
\begin{equation}\label{eqn:special1}
V_3 = x^3 + \frac{3}{16} x y^2
\end{equation}
with an additional first integral 
\begin{equation}\label{eqn:genquartic1}
{\mathit \Phi} = p_y^4 - \frac{1}{4} y^3 p_x p_y 
+ \frac{3}{4} x y^2 p_y^2 - \frac{3}{64} x^2 y^4 - \frac{1}{128} y^6.
\end{equation}
\endnumparts
Ramani~\etal~\cite{ramani1982} found the potential of degree 3, 
\numparts
\begin{equation}\label{eqn:special2}
V_3 = x^3 + \frac12 x y^2 + \frac{\sqrt{3}\,i}{18} y^3
\end{equation}
with an additional first integral
\begin{eqnarray}\label{eqn:genquartic2}
\fl \Phi = p_x p_y^3 - \frac{\sqrt{3}\,i}{2} p_y^4 + \frac12 y^3 p_x^2 
- \left(\frac32 x y^2 - \frac{\sqrt{3}\,i}{2} y^3\right) p_x p_y 
+ \left(3 x^2 y - \sqrt{3}\,i x y^2 + \frac12 y^3\right) p_y^2 
\nonumber \\
\lo + \frac12 x^3 y^3 + \frac{\sqrt{3}\,i}{8} x^2 y^4 - \frac14 x y^5 
+ \frac{5 \sqrt{3}\,i}{72} y^6
\end{eqnarray}
\endnumparts
and the potential of degree 4, 
\numparts
\begin{equation}\label{eqn:special3}
V_4 = x^4 + \frac{3}{4} x^2 y^2 + \frac{1}{8} y^4
\end{equation} 
with an additional first integral
\begin{equation}\fl \label{eqn:genquartic3}
{\mathit \Phi} = p_y^4 + \frac{1}{2} y^4 p_x^2 - 2 x y^3 p_x p_y + 
\left(3 x^2 y^2 + \frac{1}{2} y^4\right) p_y^2  + 
\frac{1}{4} x^4 y^4 + \frac{1}{4} x^2 y^6 + \frac{1}{16} y^8.
\end{equation}
\endnumparts
All of the three polynomial first integrals (\ref{eqn:genquartic1}), 
(\ref{eqn:genquartic2}), (\ref{eqn:genquartic3}) are {\em genuinely} 
quartic in the momenta. 
Here, we mean by `{\em genuinely}' that these first integrals 
cannot be reduced to first integrals of lower orders in the momenta. 
See also \cite{gramma1983} for the discoveries of these three 
integrable cases. 

Now the following question arises. 
{\em Are there any other potentials that admit a polynomial first 
integral which is genuinely quartic $($or cubic$)$ in the momenta?}
Hietarinta~\cite{hieta1983} searched integrable 2D homogeneous 
polynomial potentials of degree 5 or less with a polynomial first 
integral of order 4 or less in the momenta by means of the so-called 
direct method and concluded that there was no integrable case 
other than the known ones in the range studied. 
We extended the degree of homogeneous polynomial 
potentials to an arbitrary positive integer. 
Specifically, we investigated the existence of polynomial first 
integrals which are cubic or quartic in the momenta for 
Hamiltonian systems of the form
\begin{equation}\label{eqn:system2}
H = \frac{1}{2} (p_x^2 + p_y^2) + \alpha_0 x^k + \alpha_2 x^{k-2} y^2 
+ \cdots + \alpha_k y^k,
\end{equation}
with the degree of the potential \(k \geq 3\) 
(the cases for \(k=1,2\) are obviously integrable). 
Note that the term of \(x^{k-1} y\) in the potential vanishes 
for the removal of rotational degrees of freedom as mentioned in 
section~\ref{sec:intro}.
As a result, we obtained the following theorem.
\begin{thm}\label{thm:main}
If \(k \geq 5\), then the Hamiltonian system $(\ref{eqn:system2})$ 
admits no polynomial first integrals which are genuinely cubic or 
quartic in the momenta.
\end{thm}
Therefore, the answer to the above question is ``No.''

\section{Amplification of the statement of theorem~\ref{thm:main}}
\label{sec:state}
\noindent
In this section, we give details of the statement of 
theorem~\ref{thm:main}. 
Without loss of generality, we can assume that a polynomial first 
integral \({\mathit \Phi}\) has the following properties. 

\bigskip
\noindent
{\bf Property 1.} A polynomial first integral \({\mathit \Phi}\) is 
either even or odd in the momenta \((p_x,p_y)\). \\
{\bf Property 2.} A polynomial first integral \({\mathit \Phi}\) is 
weighted-homogeneous, i.e. \({\mathit \Phi}\) satisfies
\begin{equation}\label{eqn:weighted}
{\mathit \Phi} (\sigma^{2/(k-2)} x, \sigma^{2/(k-2)} y, 
\sigma^{k/(k-2)} p_x,\sigma^{k/(k-2)} p_y) 
= \sigma^M {\mathit \Phi}(x,y,p_x,p_y),
\end{equation}
where \(\sigma\) is an arbitrary constant and \(M\) is a constant 
called a {\em weight}. %which depends on \(k\). 
The property 1 is due to the time reflection symmetry of the system. 
The property 2 is due to the scale-invariance of the system, 
which arises from the homogeneity of potentials.
See \ref{sec:method} for more details. It is easy to check 
that all the polynomial first integrals in section~\ref{sec:known} 
satisfy the properties 1 and 2.

\subsection{Polynomial first integrals cubic in the momenta}
\noindent
From the property 1, we can put a polynomial first integral 
which is cubic in the momenta in the form
\begin{eqnarray}\label{eqn:integral3}
{\mathit \Phi} = && A_0(x,y) p_x^3 + A_1(x,y) p_x^2 p_y 
+ A_2(x,y) p_x p_y^2 + A_3(x,y) p_y^3 \nonumber \\
&& + B_0(x,y) p_x + B_1(x,y) p_y.
\end{eqnarray}
If we regard the PDE (\ref{eqn:poisson}) as an identity for the 
momenta \((p_x,p_y)\), then we have the following three sets of 
PDEs.
\begin{eqnarray}
&& \left\{
\begin{array}{l}
A_{0x} = 0 \\
A_{0y} + A_{1x} = 0 \\
A_{1y} + A_{2x} = 0 \\
A_{2y} + A_{3x} = 0 \\
A_{3y} = 0
\end{array}
\right. \label{eqn:cubic1} \\
&& \left\{
\begin{array}{l}
B_{0x} = 3 A_0 V_{x} + A_1 V_{y} \\ 
B_{0y} + B_{1x} = 2 A_1 V_{x} + 2 A_2 V_{y} \\
B_{1y} = A_2 V_{x} + 3 A_3 V_{y} 
\end{array}
\right. \label{eqn:cubic2} \\
&& \quad B_0 V_{x} + B_1 V_{y} = 0 \label{eqn:cubic3}
\end{eqnarray}
where the subscripts \(x\) and \(y\) denote partial derivatives. 
From the property 2, the polynomials \(A_i(x,y)\) are homogeneous 
of the same degree and so are the polynomials \(B_i(x,y)\). 
The PDEs~(\ref{eqn:cubic1}) have four homogeneous polynomial 
solutions, which classify the leading part of the first integral 
(\ref{eqn:integral3}) into the following four cases.

\bigskip
\noindent
{\bf Case 1.}
\({\mathit \Phi} = a_3 (y p_x - x p_y)^3 + B_0 p_x + B_1 p_y\) \\
{\bf Case 2.}
\({\mathit \Phi} = (a_2 p_x + b_2 p_y) (y p_x - x p_y)^2 
+ B_0 p_x + B_1 p_y\) \\
{\bf Case 3.}
\({\mathit \Phi} = (a_1 p_x^2 + b_1 p_x p_y + c_1 p_y^2) 
(y p_x - x p_y) + B_0 p_x + B_1 p_y\) \\
{\bf Case 4.}
\({\mathit \Phi} = a_0 p_x^3 + b_0 p_x^2 p_y + c_0 p_x p_y^2 
+ d_0 p_y^3 + B_0 p_x + B_1 p_y\)

\bigskip
\noindent
Let us next consider the PDEs~(\ref{eqn:cubic2}). 
We obtain the PDE for \(V(x,y)\),
\begin{eqnarray}\label{eqn:cubicdarboux}
&& A_2 V_{xxx}
+ (3 A_3 - 2 A_1)V_{xxy}
+ (3 A_0 - 2 A_2) V_{xyy}
+ A_1 V_{yyy} \nonumber \\
&& + 2 (A_{2x} - A_{1y}) (V_{xx} - V_{yy}) 
 + 2 (3 A_{0y} - A_{1x} - A_{2y} + 3 A_{3x}) V_{xy} \nonumber \\
&& + (3 A_{0yy} - 2 A_{1xy} + A_{2xx}) V_x
+ (A_{1yy} - 2 A_{2xy} + 3 A_{3xx}) V_y = 0
\end{eqnarray}
by using
\begin{eqnarray}
&& \partial_y^2 (3 A_0 V_{x} + A_1 V_{y}) 
- \partial_x \partial_y (2 A_1 V_{x} + 2 A_2 V_{y}) 
+ \partial_x^2 (A_2 V_{x} + 3 A_3 V_{y}) \nonumber \\
&& = \partial_y^2 B_{0x} 
- \partial_x \partial_y (B_{0y} + B_{1x}) 
+ \partial_x^2 B_{1y} \nonumber \\
&& = 0.
\end{eqnarray}
For the above each case, the PDE (\ref{eqn:cubicdarboux}) becomes

\bigskip
\noindent
{\bf Case 1}
\numparts
\begin{eqnarray}\label{eqn:cubic-pde1}
&& a_3 \{
x^2 y V_{xxx} + (2 x y^2 - x^3)V_{xxy} 
+ (y^3 - 2 x^2 y) V_{xyy} - x y^2 V_{yyy} \nonumber \\
&& \qquad + 8 x y V_{xx} + 8(y^2 - x^2) V_{xy} - 8 x y V_{yy} 
+ 12 y V_x - 12 x V_x
\} = 0,
\end{eqnarray}
{\bf Case 2}
\begin{eqnarray}\label{eqn:cubic-pde2}
&& a_2 \{
x^2 V_{xxx} + 4 x y V_{xxy} + (3 y^2 - 2 x^2) V_{xyy} 
- 2 x y V_{yyy} \nonumber \\
&& \qquad + 8 x V_{xx} +16 y V_{xy} - 8 x V_{yy} + 12 V_x
\} + \nonumber \\
&& b_2 \{
- 2 x y V_{xxx} + (3 x^2 - 2 y^2) V_{xxy} 
+ 4 x y V_{xyy} + y^2  V_{yyy} \nonumber \\
&& \qquad - 8 y V_{xx} + 16 x V_{xy} + 8 y V_{yy} +12 V_y
\} = 0,
\end{eqnarray}
{\bf Case 3}
\begin{eqnarray}\label{eqn:cubic-pde3}
&& a_1 (
2 x V_{xxy} + 3 y V_{xyy} - x V_{yyy} + 8 V_{xy}
) + \nonumber \\
&& b_1 (
- x V_{xxx} - 2 y V_{xxy} + 2 x V_{xyy} + y V_{yyy} 
- 4 V_{xx} + 4 V_{yy}
)+ \nonumber \\
&& c_1 (
y V_{xxx} - 3 x V_{xxy} - 2 y V_{xyy} - 8 V_{xy}) = 0,
\end{eqnarray}
{\bf Case 4}
\begin{equation}\label{eqn:cubic-pde4}
a_0 (3 V_{xyy}) + b_0 (-2 V_{xxy} + V_{yyy}) 
+ c_0 (V_{xxx} - 2 V_{xyy}) + d_0 (3 V_{xxy}) = 0.
\end{equation}
\endnumparts
The potential must satisfy the two PDEs (\ref{eqn:cubicdarboux}) 
and (\ref{eqn:cubic3}). 
They are transformed into identities for \((x,y)\) by the 
substitution of the potential (\ref{eqn:potential}). 
Therefore we finally obtain two sets of simultaneous algebraic 
equations for the coefficients of the potential and of the first 
integral. 
We searched their solutions for \(k \geq 3\) under the condition 
that the leading part of the first integral does not vanish, i.e. 
\(a_3 \neq 0\), 
\((a_2,b_2) \neq (0,0)\), 
\((a_1,b_1,c_1) \neq (0,0,0)\), or 
\((a_0,b_0,c_0,d_0) \neq (0,0,0,0)\). 
Table~\ref{tab:order3} shows the results for the four cases.
\begin{table}
\caption{Integrable potentials with a first integral cubic 
in the momenta.}
\label{tab:order3}
\renewcommand{\arraystretch}{1.00}
\begin{indented}
\item[]
\begin{tabular}{@{}cll}
\br
Case & Potential & First integral \\
\mr
1 & \(V_k = r^k = (x^2 + y^2)^{k/2},\quad k=\mbox{even}\)
& \({\mathit \Phi} = (y p_x - x p_y)^3\) \\
2 & \(V_k \equiv 0\) \\
3 & \(V_k = r^k = (x^2 + y^2)^{k/2},\quad k=\mbox{even}\) 
& \({\mathit \Phi} = (y p_x - x p_y) H\) \\
4 & \(V_k = x^k\)
& \({\mathit \Phi} = b_0 p_y (p_x^2 + 2 x^k) + d_0 p_y^3 \) \\
\br
\end{tabular}
\end{indented}
\end{table}

\subsection{Polynomial first integrals quartic in the momenta}
\noindent
From the property 1, we can put a polynomial first integral which 
is quartic in the momenta in the form
\begin{eqnarray}\label{eqn:integral4}
\fl {\mathit \Phi} = A_0(x,y) p_x^4 + A_1(x,y) p_x^3 p_y 
+ A_2(x,y) p_x^2 p_y^2 + A_3(x,y) p_x p_y^3 + A_4(x,y) p_y^4 
\nonumber \\
\lo + B_0(x,y) p_x^2 + B_1(x,y) p_x p_y + B_2(x,y) p_y^2 + C_0(x,y). 
\end{eqnarray}
If we regard the PDE (\ref{eqn:poisson}) as an identity for the 
momenta \((p_x,p_y)\), then we obtain the following three sets 
of PDEs.
\begin{eqnarray}
&& \left\{
\begin{array}{l}
A_{0x} = 0 \\ 
A_{0y} + A_{1x} = 0 \\ 
A_{1y} + A_{2x} = 0 \\
A_{2y} + A_{3x} = 0 \\ 
A_{3y} + A_{4x} = 0 \\ 
A_{4y} = 0
\end{array}
\right. \label{eqn:quartic1} \\
&& \left\{
\begin{array}{l}
B_{0x} = 4 A_0 V_x + A_1 V_y \\ 
B_{0y} + B_{1x} = 3 A_1 V_x + 2 A_2 V_y \\
B_{1y} + B_{2x} = 2 A_2 V_x + 3 A_3 V_y \\ 
B_{2y} = A_3 V_x + 4 A_4 V_y
\end{array}
\right. \label{eqn:quartic2} \\
&& \left\{
\begin{array}{l}
C_{0x} = 2 B_0 V_x + B_1 V_y \\ 
C_{0y} = B_1 V_x + 2 B_2 V_y
\end{array}
\right. \label{eqn:quartic3}
\end{eqnarray}
From the property 2, the polynomials \(A_i(x,y)\) are homogeneous 
of the same degree and so are the polynomials \(B_i(x,y)\) and 
\(C_0(x,y)\). 
The PDEs~(\ref{eqn:quartic1}) have five homogeneous polynomial 
solutions, which classify the leading part of the first integral 
(\ref{eqn:integral4}) into the following five cases.

\bigskip
\noindent
{\bf Case 1.}
\({\mathit \Phi} = a_4 (y p_x - x p_y)^4 + 
B_0 p_x^2 + B_1 p_x p_y + B_2 p_y^2 + C_0\) \\
{\bf Case 2.}
\({\mathit \Phi} = (a_3 p_x + b_3 p_y) (y p_x - x p_y)^3 + 
B_0 p_x^2 + B_1 p_x p_y + B_2 p_y^2 + C_0\) \\
{\bf Case 3.}
\({\mathit \Phi} = (a_2 p_x^2 + b_2 p_x p_y + c_2 p_y^2) 
(y p_x - x p_y)^2 + B_0 p_x^2 + B_1 p_x p_y + B_2 p_y^2 + C_0\) \\
{\bf Case 4.}
\({\mathit \Phi} = (a_1 p_x^3 + b_1 p_x^2 p_y + c_1 p_x p_y^2 + 
d_1 p_y^3)(y p_x - x p_y) +
B_0 p_x^2 + B_1 p_x p_y + B_2 p_y^2 + C_0\) \\
{\bf Case 5.}
\({\mathit \Phi} = a_0 p_x^4 + b_0 p_x^3 p_y + c_0 p_x^2 p_y^2 + 
d_0 p_x p_y^3 + e_0 p_y^4 + 
B_0 p_x^2 + B_1 p_x p_y + B_2 p_y^2 + C_0\)

\bigskip
\noindent
Let us next consider the PDEs~(\ref{eqn:quartic2}). 
We obtain the PDE for \(V(x,y)\), 
\begin{eqnarray}\label{eqn:quarticdarboux1}
\fl A_3 V_{xxxx}
- 2 (A_2 - 2 A_4) V_{xxxy}
+ 3 (A_1 - A_3) V_{xxyy} 
- 2 (2 A_0 - A_2) V_{xyyy}
- A_1 V_{yyyy} \nonumber \\
\fl - \, (2 A_{2y} - 3 A_{3x}) V_{xxx}
+ (6 A_{1y} - 4 A_{2x} - 3 A_{3x} + 12 A_{4x}) V_{xxy} 
\nonumber \\
\fl - \, (12 A_{0y} - 3 A_{1x} - 4 A_{2y} + 6 A_{3x}) V_{xyy} 
- (3 A_{1y} - 2 A_{2x}) V_{yyy} \nonumber \\
\fl + \, (3 A_{1yy} - 4 A_{2xy} + 3 A_{3xx}) (V_{xx} - V_{yy}) 
\nonumber \\
\fl - \, (12 A_{0yy} - 6 A_{1xy} + 2 A_{2xx} - 2 A_{2yy} 
+ 6 A_{3xy} - 12 A_{4xx}) V_{xy} \nonumber \\
\fl - \, (4 A_{0yyy} - 3 A_{1xyy} + 2 A_{2xxy} - A_{3xxx}) V_x 
\nonumber \\
\fl - \, (A_{1yyy} - 2 A_{2xyy} + 3 A_{3xxy} - 4 A_{4xxx}) V_y = 0
\end{eqnarray}
by using
\begin{eqnarray}
\fl \partial_y^3 (4 A_0 V_x + A_1 V_y) 
- \partial_x \partial_y^2 (3 A_1 V_x + 2 A_2 V_y) 
+ \partial_x^2 \partial_y (2 A_2 V_x + 3 A_3 V_y) 
- \partial_x^3 (A_3 V_x + 4 A_4 V_y) \nonumber \\
= \partial_y^3 B_{0x} 
- \partial_x \partial_y^2 (B_{0y} + B_{1x}) 
+ \partial_x^2 \partial_y (B_{1y} + B_{2x}) 
- \partial_x^3 B_{2y} \nonumber \\
=  0.
\end{eqnarray}
For the above each case, the PDE (\ref{eqn:quarticdarboux1}) 
becomes 

\bigskip
\noindent
{\bf Case 1}
\numparts
\begin{eqnarray}\label{eqn:quartic-pde1}
\fl a_4 \{
x^3 y V_{xxxx} - (x^4 - 3 x^2 y^2) V_{xxxy} 
- 3(x^3 y - x y^3) V_{xxyy} 
+ (y^4 - 3 x^2 y^2) V_{xyyy} - x y^3 V_{yyyy} \nonumber \\
\lo 
+ 15 x^2 y V_{xxx} - 15(x^3 - 2 x y^2) V_{xxy} 
+ 15(y^3 - 2 x^2 y) V_{xyy}
- 15 x y^2 V_{yyy} \nonumber \\
\lo 
+ 60 x y V_{xx} - 60(x^2 - y^2) V_{xy} - 60 x y V_{yy} 
+ 60 y V_x - 60 x V_y \} = 0,
\end{eqnarray}
{\bf Case 2}
\begin{eqnarray}\label{eqn:quartic-pde2}
\fl a_3 \{
x^3 V_{xxxx} + 6 x^2 y V_{xxxy} - 3(x^3 - 3 x y^2) V_{xxyy}
- 2(3 x^2 y - 2 y^3) V_{xyyy} - 3 x y^2 V_{yyyy} \nonumber \\
\lo 
+ 15 x^2 V_{xxx} + 60 x y V_{xxy} - 15(2 x^2 - 3 y^2) V_{xyy} 
- 30 x y V_{yyy} \nonumber \\
\lo 
+ 60 x V_{xx} +120 y V_{xy} - 60 x V_{yy} + 60 V_x
\} + \nonumber  \\
\fl b_3 \{
- 3 x^2 y V_{xxxx} + 2(2 x^3 - 3 x y^2) V_{xxxy} 
+ 3(3 x^2 y - y^3) V_{xxyy}
+ 6 x y^2 V_{xyyy} + y^3 V_{yyyy} \nonumber \\
\lo 
- 30 x y V_{xxx} + 15(3 x^2 - 2 y^2) V_{xxy} + 60 x y V_{xyy} 
+ 15 y^2 V_{yyy}
\nonumber \\
\lo 
- 60 y V_{xx}-120 x V_{xy} + 60 y V_{yy} + 60 V_y
\} = 0, 
\end{eqnarray}
{\bf Case 3}
\begin{eqnarray}\label{eqn:quartic-pde3}
\fl a_2 \{
2 x^2 V_{xxxy} + 6 x y V_{xxyy} - 2(x^2 - 2 y^2) V_{xyyy} 
- 2 x y V_{yyyy} \nonumber \\
\lo 
+ 20 x V_{xxy} + 30 y V_{xyy} - 10 x V_{yyy} + 40 V_{xy}\} + 
\nonumber \\
\fl b_2 \{
- x^2 V_{xxxx} - 4 x y V_{xxxy} + 3(x^2 - y^2) V_{xxyy}
+ 4 x y V_{xyyy} + y^2 V_{yyyy} \nonumber \\
\lo 
- 10 x V_{xxx} -  20 y V_{xxy} + 20 x V_{xyy} + 10 y V_{yyy} 
- 20 V_{xx} + 20 V_{yy}\} + \nonumber \\
\fl c_2 \{
2 x y V_{xxxx} - 2(2 x^2 - y^2) V_{xxxy} - 6 x y V_{xxyy} 
- 2 y^2 V_{xyyy}\nonumber \\
\lo 
+ 10 y V_{xxx} + 30 x V_{xxy} - 20 y V_{xyy} - 40 V_{xy}\} = 0,
\end{eqnarray}
{\bf Case 4}
\begin{eqnarray}
a_1 (
3 x V_{xxyy} + 4 y V_{xyyy} - x V_{yyyy} + 15 V_{xyy}) + 
\nonumber \\
b_1 (
-2 x V_{xxxy} - 3 y V_{xxyy} + 2 x V_{xyyy} 
+ y V_{yyyy} - 10 V_{xxy} + 5 V_{yyy}) + \nonumber \\
c_1 (
x V_{xxxx} + 2 y V_{xxxy} - 3 x V_{xxyy} 
- 2 y V_{xyyy} + 5 V_{xxx} - 10 V_{xyy}) + \nonumber \\
d_1 (
- y V_{xxxx} + 4 x V_{xxxy} + 3 y V_{xxyy} + 15 V_{xxy}) = 0,
\label{eqn:quartic-pde4}
\end{eqnarray}
{\bf Case 5}
\begin{eqnarray}
& a_0 (4 V_{xyyy}) 
+ b_0 (-3 V_{xxyy} + V_{yyyy}) 
+ c_0 (2 V_{xxxy} - 2 V_{xyyy}) \nonumber \\ 
& + d_0 (-V_{xxxx}+3 V_{xxyy}) + e_0 (-4 V_{xxxy}) = 0.
\label{eqn:quartic-pde5}
\end{eqnarray}
\endnumparts
We also obtain the PDEs for \(V(x,y)\),
\begin{equation}\fl \label{eqn:quarticdarboux2}
B_1 (V_{xx} - V_{yy}) + 2(B_2 - B_0) V_{xy} + 
(B_{1x} - 2 B_{0y}) V_x + (2 B_{2x} - B_{1y}) V_y = 0
\end{equation}
from (\ref{eqn:quartic3}) by using
\begin{equation}
\partial_y (2 B_0 V_x + B_1 V_y) 
- \partial_x (B_1 V_x + 2 B_2 V_y) 
= \partial_y C_{0x} - \partial_x C_{0y} = 0. 
\end{equation}
The potential must satisfy the two PDEs (\ref{eqn:quarticdarboux1}) 
and (\ref{eqn:quarticdarboux2}). 
They are transformed into identities for \((x,y)\) by the substitution 
of the potential (\ref{eqn:potential}). 
Therefore, we finally obtain two sets of simultaneous algebraic 
equations for the coefficients of the potential and of the first 
integral. 
We searched their solutions for \(k \geq 3\) under the condition 
that the leading part of the first integral does not vanish, i.e. 
\(a_4 \neq 0\), \((a_3,b_3) \neq (0,0)\), 
\((a_2,b_2,c_2) \neq (0,0,0)\), 
\((a_1,b_1,c_1,d_1) \neq (0,0,0,0)\), 
or \((a_0,b_0,c_0,d_0,e_0) \neq (0,0,0,0,0)\).
Table~\ref{tab:order4} shows the results obtained from the five cases, 
which hold for \(k \geq 3\) except that there exist three exceptional 
potentials, (\ref{eqn:special1}), (\ref{eqn:special2}), 
(\ref{eqn:special3}), in the case 5 for \(k=3,4\). 
\begin{table}
\caption{Integrable potentials with a first integral quartic 
in the momenta.}
\label{tab:order4}
\renewcommand{\arraystretch}{1.00}
\begin{indented}
\item[]
\begin{tabular}{@{}cll}
\br
Case & Potential & First integral \\
\mr
1 & \(V_k = r^k = (x^2 + y^2)^{k/2},\quad k=\mbox{even}\)
& \({\mathit \Phi} = (y p_x - x p_y)^4\) \\
2 & \(V_k \equiv 0\) \\
3 & \(V_k = r^k = (x^2 + y^2)^{k/2},\quad k=\mbox{even}\) 
& \({\mathit \Phi} = (y p_x - x p_y)^2 H\) \\
& \(
V_k = \frac{1}{r}
\left[\left(\frac{r+x}{2}\right)^{k+1} + (-1)^k 
\left(\frac{r-x}{2}\right)^{k+1}\right]
\) 
& \(
{\mathit \Phi} = \left(p_y (y p_x - x p_y) 
+ \frac{1}{2}y^2 V_{k-1}\right)^2
\) \\
4 & \(
V_k = \frac{1}{r}
\left[\left(\frac{r+x}{2}\right)^{k+1} + (-1)^k 
\left(\frac{r-x}{2}\right)^{k+1}\right]
\) 
& \(
{\mathit \Phi} = \left(p_y (y p_x - x p_y) 
+ \frac{1}{2}y^2 V_{k-1}\right) H
\) \\
5 & \(V_k = A x^k + B y^k\)
& \({\mathit \Phi} = a_0 (p_x^2 + 2 A x^k)^2 
+ e_0 (p_y^2 + 2 B y^k)^2\) \\
\br
\end{tabular}
\end{indented}
\end{table}

\bigskip
\noindent
All the potentials in tables~\ref{tab:order3} and \ref{tab:order4} 
are separable ones given in section~\ref{sec:known} 
and the corresponding first integrals are {\em apparently} cubic and 
quartic in the momenta, i.e. there are no new integrable cases. 
This is what the statement of theorem~\ref{thm:main} means.

\section{Details of the computations}\label{sec:detail}
\noindent
All of the three polynomial first integrals (\ref{eqn:genquartic1}), 
(\ref{eqn:genquartic2}), (\ref{eqn:genquartic3}), which are 
{\em genuinely} quartic in the momenta, fall into the case 5. 
So, it is quite natural to expect that new integrable potentials, 
if any, would have first integrals which belong to the case 5. 
For this reason, we take the case 5 as an example to show the details 
of the computations performed in this study.
We first exclude the square of the Hamiltonian by considering 
\(\Phi - 2c_0 H^2\) instead of \(\Phi\) itself, i.e. we put \(c_0=0\) 
from the beginning. 
The PDEs~(\ref{eqn:quartic2}) for this case become
\begin{equation}\label{eqn:quartic2-5}
\left\{
\begin{array}{l}
B_{0x} = 4 a_0 V_x + b_0 V_y \\ 
B_{0y} + B_{1x} = 3 b_0 V_x \\
B_{1y} + B_{2x} = 3 d_0 V_y \\ 
B_{2y} = d_0 V_x + 4 e_0 V_y
\end{array}
\right.
\end{equation}
We obtain the expressions of \(B_0, B_1, B_2\) 
by integrating (\ref{eqn:quartic2-5}) as follows (note that \(\alpha_1=0\)).
\begin{eqnarray}
\fl B_0 = \sum_{j=0}^{k-1} \frac{4(k-j)\alpha_j a_0 
+ (j+1)\alpha_{j+1}b_0}{k-j}x^{k-j}y^j+r_0 y^k, \\
\fl B_1 =  r_1 x^k - k r_2 x^{k-1}y \nonumber \\ 
\fl \qquad + \sum_{j=2}^k \left[
\left\{3 \alpha_j 
- \frac{(k-j+1)(k-j+2)}{j(j-1)}\alpha_{j-2}\right\}d_0 
- \frac{4(k-j+1)}{j}\alpha_{j-1} e_0
\right] x^{k-j}y^j, \\
\fl B_2 = r_2 x^k 
+ \sum_{j=1}^k \frac{(k-j+1)\alpha_{j-1}d_0 
+ 4j\alpha_j e_0}{j}x^{k-j}y^j,
\end{eqnarray}
where
\begin{eqnarray}
\fl r_0 = \frac{3(k-1)(k-2)\alpha_{k-1}b_0 
+ \{6\alpha_{k-3}-3(k-1)(k-2)\alpha_{k-1}\}d_0 
+ 8(k-2)\alpha_{k-2}e_0}{k(k-1)(k-2)}, \\
\fl r_1 = \frac{8(k-2)\alpha_2 a_0 + 6\alpha_3 b_0}{k(k-1)(k-2)}, \\
\fl r_2 = \frac{\{3k(k-1)\alpha_0 - 2\alpha_2\}b_0}{k(k-1)}.
\end{eqnarray}
The PDE (\ref{eqn:quarticdarboux1}), or 
(\ref{eqn:quartic-pde5}) becomes
\begin{equation}
4 a_0 V_{xyyy} 
- b_0 (3 V_{xxyy} - V_{yyyy}) 
-  d_0 (V_{xxxx} - 3 V_{xxyy}) -4 e_0 V_{xxxy} = 0.
\label{eqn:pde5}
\end{equation}
From (\ref{eqn:pde5}) and (\ref{eqn:quarticdarboux2}) 
with \(B_0,B_1,B_2\) given above, we obtain two sets of 
simultaneous algebraic equations 
for the coefficients of the potential, \(\alpha_j\), and 
of the leading part of the first integral, \(a_0,b_0,d_0,e_0\). 
The simultaneous algebraic equations obtained from (\ref{eqn:pde5}) 
consist of the following \(k-3\) equations.
\begin{eqnarray}\label{eqn:rec4(v)}
\fl 4(j-1)j(j+1)(k-j-1)\alpha_{j+1} a_0 \nonumber \\
\fl + \, j(j-1)\{(j+1)(j+2)\alpha_{j+2} - 3(k-j)(k-j-1)\alpha_j\} b_0 
\nonumber \\
\fl + \, (k-j)(k-j-1)\{3j(j-1)\alpha_j 
- (k-j+1)(k-j+2)\alpha_{j-2}\} d_0 \nonumber \\
\fl - \, 4(j-1)(k-j-1)(k-j)(k-j+1)\alpha_{j-1} e_0 = 0 
\quad (j=2,3,\ldots,k-2)
\end{eqnarray}
which can be regarded as a recurrence relation for \(\alpha_j\). 
The simultaneous algebraic equations obtained from 
(\ref{eqn:quarticdarboux2}) consist of \(2k-1\) equations 
of the form
\begin{equation}
\left\{
\begin{array}{l}
M_{11} b_0 = 0 \\
M_{21} b_0 + M_{22} a_0 = 0 \\
M_{31} b_0 + M_{32} a_0 + M_{33} d_0 = 0 \\
M_{41} b_0 + M_{42} a_0 + M_{43} d_0 + M_{44} e_0 = 0 \\
M_{51} b_0 + M_{52} a_0 + M_{53} d_0 + M_{54} e_0 = 0 \\
M_{61} b_0 + M_{62} a_0 + M_{63} d_0 + M_{64} e_0 = 0 \\
\quad \cdots\cdots\cdots\cdots\cdots\cdots\cdots\cdots\cdots\\
M_{2k-1,1} b_0 + M_{2k-1,2} a_0 + M_{2k-1,3} d_0 
+ M_{2k-1,4} e_0 = 0
\end{array}
\right.
\label{eqn:sim4(v)}
\end{equation}
where the first four equations are given by the following.
\begin{eqnarray*}
\fl M_{11} = 
\frac{\{3k(2k-1)\alpha_0 - 2\alpha_2\}
\{k(k-1)\alpha_0 - 2\alpha_2\}}{k(k-1)}, \\
\fl M_{21} =
-\frac{6(7k-6)\alpha_0\alpha_3}{k-2} + 
\frac{12(7k-6)\alpha_2\alpha_3}{k(k-1)(k-2)}, \\
\fl M_{22} = - \frac{16(3k-2)\{k(k-1)\alpha_0 
- 2\alpha_2\}\alpha_2}{k(k-1)}, \\
\fl M_{31} = 6(k-1)(2k-3)\alpha_0\alpha_2 
- \frac{2(17k^2-28k+6)\alpha_2^2}{k(k-1)} + 
\frac{18(7k-6)\alpha_3^2}{k(k-1)(k-2)} \\ 
\lo - \frac{12(7k^2-22k+18)\alpha_0\alpha_4}{(k-2)(k-3)} + 
\frac{48(2k^2-4k+3)\alpha_2\alpha_4}{k(k-1)(k-2)(k-3)}, \\
\fl M_{32} = -\frac{12(2k-3)(3k-4)\alpha_0\alpha_3}{k-2} +
\frac{24(2k-3)(5k-4)\alpha_2\alpha_3}{k(k-1)(k-2)}, \\
\fl M_{33} = 4k(2k-3)\alpha_0 \alpha_2, \\
\fl M_{41} = 6(k-2)(2k-3)\alpha_0\alpha_3 - 
\frac{4(29k^2-44k+6)\alpha_2\alpha_3}{k(k-1)} + 
\frac{48(7k-12)\alpha_3\alpha_4}{k(k-2)(k-3)} \\
\lo - \frac{20(7k^2-31k+36)\alpha_0\alpha_5}{(k-3)(k-4)} + 
\frac{40(5k^2-11k+12)\alpha_2\alpha_5}{k(k-1)(k-3)(k-4)}, \\
\fl M_{42} = - \frac{16(k-2)(3k-4)\alpha_2^2}{k-1} + 
\frac{144\alpha_3^2}{k-2} 
- \frac{96(k-2)^2\alpha_0\alpha_4}{k-3} + 
\frac{384(k-2)\alpha_2\alpha_4}{k(k-3)}, \\
\fl M_{43} = 12 k(k-2)\alpha_0\alpha_3, \quad 
M_{44} = 32(k-2) \alpha_2^2
\end{eqnarray*}
Now, what we have to do is to find solutions of (\ref{eqn:rec4(v)}) 
and (\ref{eqn:sim4(v)}) under the condition 
\((b_0,a_0,d_0,e_0) \neq (0,0,0,0)\). 
Applying this condition to the first four equations of 
(\ref{eqn:sim4(v)}), we see that the product 
\(M_{11} M_{22} M_{33} M_{44}\) must vanish. 
We therefore have the following possibilities for the relation 
between \(\alpha_0\) and \(\alpha_2\). 
\begin{equation}\fl \label{eqn:relation}
\mbox{(i) }\alpha_0 = 0 \quad
\mbox{(ii) }\alpha_2 = 0 \quad
\mbox{(iii) }\alpha_2 = \frac{k(k-1)}{2} \alpha_0 \quad
\mbox{(iv) }\alpha_2 = \frac{3k(2k-1)}{2} \alpha_0
\end{equation}
Once the relation between \(\alpha_0\) and \(\alpha_2\) is given, 
the computations to follow are quite straightforward. 
Let us move on to the details of each case.

\bigskip
\noindent
(i) When \(\alpha_0=0\), the first and the second equations of 
(\ref{eqn:sim4(v)}) become
\begin{equation}
\left\{
\begin{array}{l}
\displaystyle{
\frac{4\alpha_2^2}{k(k-1)}b_0 = 0} \\
\displaystyle{
\frac{12(7k-6)\alpha_2 \alpha_3}{k(k-1)(k-2)} b_0 + 
\frac{32(3k-2)\alpha_2^2}{k(k-1)} a_0 = 0}
\end{array}
\right.
\end{equation}
If we assume that \(\alpha_2 \neq 0\) 
then we see that \(b_0 = a_0 = 0\), and then 
we can show from (\ref{eqn:rec4(v)}) with \(j=2,3\) 
that \(d_0=e_0=0\). 
This contradicts the condition \((b_0,a_0,d_0,e_0) \neq (0,0,0,0)\). 
Therefore, \(\alpha_2\) must vanish. 
Then the third equation of (\ref{eqn:sim4(v)}) becomes
\begin{equation}
\frac{18(7k-6)\alpha_3^2}{k(k-1)(k-2)} b_0 = 0.
\end{equation}
Let us assume that \(\alpha_3 \neq 0\). 
Then, \(b_0=0\) and we can show from (\ref{eqn:rec4(v)}) 
with \(j=2,3,4\) that \(a_0=d_0=e_0=0\). 
This contradicts the condition \((b_0,a_0,d_0,e_0) \neq (0,0,0,0)\). 
Therefore, \(\alpha_3\) must vanish.
Let us now suppose that we have proved that 
\(\alpha_0 = \alpha_1 = \cdots = \alpha_j = 0\) \((j \geq 3)\). 
Then it follows from (\ref{eqn:rec4(v)}) 
that \(\alpha_{j+1}\) must vanish, up to \(j = k-4\), under the 
condition \((b_0,a_0,d_0,e_0) \neq (0,0,0,0)\). 
Now, we have \(\alpha_0=\alpha_1=\cdots=\alpha_{k-3}=0\). 
If we assume that \(\alpha_{k-2} \neq 0\) 
then we have \(b_0 = a_0 = d_0 = 0\) 
from (\ref{eqn:rec4(v)}) with \(j=k-4,k-3,k-2\). 
Then the \((2k-4)\)-th equation of (\ref{eqn:sim4(v)}) becomes
\begin{equation}
\frac{16(k-2)(3k-4)\alpha_{k-2}^2}{k-1} e_0 = 0
\end{equation}
which shows that \(e_0=0\). 
This contradicts the condition \((b_0,a_0,d_0,e_0) \neq (0,0,0,0)\). 
Therefore, \(\alpha_{k-2}\) must vanish. 
Let us assume that \(\alpha_{k-1} \neq 0\). 
Then we have \(b_0 = a_0 = 0\) from (\ref{eqn:rec4(v)}) 
with \(j=k-3,k-2\). 
The \((2k-3)\)-rd and the \((2k-2)\)-nd equations of 
(\ref{eqn:sim4(v)}) become
\begin{equation}
\left\{
\begin{array}{l}
- 3(k-1)(2k-3)\alpha_{k-1}^2 d_0 = 0 \\
\displaystyle{
- 6(k-1)(2k-1)\alpha_{k-1}\alpha_k d_0 + 
\frac{8(k-1)(3k-1)\alpha_{k-1}^2}{k} e_0 = 0}
\end{array}
\right.
\end{equation}
which show that \(d_0 = e_0 = 0\). 
This contradicts the condition \((b_0,a_0,d_0,e_0) \neq (0,0,0,0)\). 
Therefore, \(\alpha_{k-1}\) must vanish. 
It has shown that \(\alpha_0 = \alpha_1 = \cdots = \alpha_{k-1} = 0\). 
Then the \((2k-3)\)-rd and the \((2k-1)\)-st equations of 
(\ref{eqn:sim4(v)}) become
\begin{equation}
\left\{
\begin{array}{l}
\displaystyle{
\frac{1}{2} k^2 (k-1)(2k-3)\alpha_k^2 b_0 = 0} \\
- 3k(2k-1)\alpha_k^2 d_0 = 0
\end{array}
\right.
\end{equation}
The only possible solution is \(b_0 = d_0 = 0\) 
with \(\alpha_k \neq 0\). 
Therefore, we obtain 
\begin{equation}\label{eqn:yk}
V_k = y^k.
\end{equation}
If we assume that \(\alpha_0=0\) for the other three cases 
(ii), (iii), (iv), then we only obtain the potential (\ref{eqn:yk}). 
Therefore, we assume that \(\alpha_0 \neq 0\) for the cases 
(ii), (iii), (iv). Let us put \(\alpha_0=1\). 

\bigskip
\noindent
(ii) When \(\alpha_2=0\) and \(\alpha_0=1\), 
we can see from the first equation of (\ref{eqn:sim4(v)}) 
that \(b_0=0\). 
Then the third equation of (\ref{eqn:sim4(v)}) becomes
\begin{equation}
- \frac{12(2k-3)(3k-4)\alpha_3}{k-2} a_0 = 0, 
\end{equation}
which shows that \(\alpha_3 a_0 = 0\). 
Then the recurrence relation (\ref{eqn:rec4(v)}) gives 
\begin{equation}
 - k(k-1)(k-2)(k-3) d_0 = 0 \quad (j=2)
\end{equation}
from which we obtain \(d_0 = 0\). 
Let us now suppose that we have proved that
\(\alpha_1 = \alpha_2 = \cdots = \alpha_j = 0\) (\(j \geq 2\)). 
Then it follows from (\ref{eqn:rec4(v)}) 
that \(\alpha_{j+1}\) must vanish, up to \(j = k-4\), under the 
condition \((a_0,e_0) \neq (0,0)\). 
Now, we have \(\alpha_1 = \alpha_2 = \cdots = \alpha_{k-3} = 0\). 
If we assume that \(\alpha_{k-2} \neq 0\) then 
we can see from (\ref{eqn:rec4(v)}) with \(j=k-3\) 
that \(a_0=0\).
Then the \(k\)th equation of (\ref{eqn:sim4(v)}) reads
\begin{equation}
- \frac{8k(k+2)\alpha_{k-2}}{k-1} e_0 = 0, 
\end{equation}
which shows that \(e_0=0\) since \(\alpha_{k-2} \neq 0\). 
This contradicts the condition \((b_0,a_0,d_0,e_0) \neq (0,0,0,0)\). 
Therefore, \(\alpha_{k-2}\) must vanish. 
Next, let us assume that \(\alpha_{k-1} \neq 0\). 
Then we can see from (\ref{eqn:rec4(v)}) with \(j=k-2\) 
that \(a_0=0\). 
Then the \((k+1)\)-st equation of (\ref{eqn:sim4(v)}) becomes
\begin{equation}
- 4(k-1)\alpha_{k-1} e_0 = 0, 
\end{equation}
which shows that \(e_0=0\) since \(\alpha_{k-1} \neq 0\).
This contradicts the condition \((b_0,a_0,d_0,e_0) \neq (0,0,0,0)\). 
Therefore, \(\alpha_{k-1}\) must vanish. 
Then it has shown that 
\(\alpha_1 = \alpha_2 = \cdots = \alpha_{k-1} = 0\). 
Thus, we obtain 
\begin{equation}\label{eqn:xkyk}
V_k = x^k + \alpha_k y^k.
\end{equation}

\bigskip
\noindent
(iii) When \(\alpha_2=(k(k-1)/2)\alpha_0\) and \(\alpha_0=1\), 
the third and the fourth equations of (\ref{eqn:sim4(v)}) yield
\begin{eqnarray}
&& d_0 = - \frac{M_{32} a_0 + M_{31} b_0}{M_{33}}, \\
&& e_0 = - \frac{1}{M_{44}} \left\{
\left(M_{42} + \frac{M_{43}M_{32}}{M_{33}} \right) a_0 + 
\left(M_{41} + \frac{M_{43}M_{31}}{M_{33}} \right) b_0 \right\}. 
\end{eqnarray}
Then the condition \((b_0,a_0,d_0,e_0) \neq (0,0,0,0)\) reads 
\((b_0,a_0) \neq (0,0)\). We obtain
\begin{equation}\fl \label{eqn:alphaj}
\alpha_j = {k \choose j} \left[
\frac{(\sin \varphi_0)^j}{(\cos \varphi_0)^{j-2}} 
\pm (-1)^{k-j} 
\frac{(\cos \varphi_0)^j}{(\sin \varphi_0)^{j-2}} \right]; 
\quad 
\tan 2\varphi_0 = - \frac{2}{\alpha_3} {k \choose 3}
\end{equation}
from the recurrence relation (\ref{eqn:rec4(v)}) 
under the condition \((b_0,a_0) \neq (0,0)\). 
Here, we take the positive sign for even \(k\) and the negative sign 
for odd \(k\). 
The potential with (\ref{eqn:alphaj}) is transformed 
into the separable form
\begin{equation}
V_k = (\sec \varphi_0)^{k-2} x^k \pm 
(\mbox{cosec}\, \varphi_0)^{k-2} y^k 
\end{equation}
by the rotation of coordinates defined by
\begin{equation}\label{eqn:rotation}
x \to x \cos \varphi_0 - y \sin \varphi_0, \quad
y \to x \sin \varphi_0 + y \cos \varphi_0.
\end{equation}

\bigskip
\noindent
(iv) When \(\alpha_2=(3k(2k-1)/2)\alpha_0\) and \(\alpha_0=1\), 
the second, the third, and the fourth equations of (\ref{eqn:sim4(v)}) 
yield
\begin{eqnarray}
&& a_0 = - \frac{M_{21}}{M_{22}} b_0, \\
&& d_0 = - \frac{1}{M_{33}}
\left(M_{31} - M_{32} \frac{M_{21}}{M_{22}} \right) b_0, \\
&& e_0 = - \frac{1}{M_{44}} 
\left\{M_{41} - M_{42} \frac{M_{21}}{M_{22}} - M_{43} 
\frac{1}{M_{33}}
\left(M_{31} - M_{32} \frac{M_{21}}{M_{22}}\right)\right\} b_0.
\end{eqnarray}
Then the condition \((b_0,a_0,d_0,e_0) \neq (0,0,0,0)\) reads 
\(b_0 \neq 0\). 
Let us put \(b_0=1\). Then, from the recurrence relation 
(\ref{eqn:rec4(v)}), we can express 
\(\alpha_4, \alpha_5, \ldots, \alpha_k\) as polynomials of 
\(\alpha_3\) whose coefficients are rational functions of \(k\). 
Therefore, the rest of equations (\ref{eqn:sim4(v)}), 
the number of which is \(2k-5\), yield algebraic equations 
for \(\alpha_3\). 
Now, the problem is whether the \(2k-5\) algebraic equations 
for \(\alpha_3\) have common solutions or not. 
The two algebraic equations of the lowest degrees, which are 
obtained from the fifth and the sixth equations of (\ref{eqn:sim4(v)}), 
are explicitly given by
\begin{eqnarray}
&& - \frac{9(k - 2)k^2(k + 2)(2k - 1)(5k - 2) G_1(k)}
{4(k - 1) S(k)^2} \nonumber \\
&& + \frac{3(k + 2)(5k - 2)(7k - 6) G_2(k) \alpha_3^2}
{4(k - 2)(k - 1)(2k - 1)(3k - 2) R(k) S(k)^2} \nonumber \\
&& + 
\frac{2(k + 2)(5k - 2)(7k - 6) G_3(k) \alpha_3^4}
{(k - 2)^3(k - 1)k^2(2k - 1)^3(3k - 2)^3 R(k) S(k)^2} = 0
\label{eqn:eqn1}
\end{eqnarray}
and
\begin{eqnarray}
&& - \frac{9(k - 3)k(k + 2)(5k - 2) G_4(k) \alpha_3}
{20(k - 1)(3k - 2) R(k) S(k)^2} \nonumber \\
&& + \frac{3(k - 3)(k + 2)(5k - 2)(7k - 6) G_5(k) \alpha_3^3}
{20(k - 2)^2(k - 1)k(2k - 1)^2(3k - 2)^2 R(k) S(k)^2} \nonumber \\
&& + 
\frac{(k - 3)(k + 2)(5k - 2)(7k - 6) G_6(k) \alpha_3^5}
{5(k - 2)^4(k - 1)k^3(2k - 1)^4(3k - 2)^4 R(k) S(k)^2} = 0, 
\label{eqn:eqn2}
\end{eqnarray}
where
\begin{eqnarray*}
\fl R(k) = 74 - 151k + 95k^2, \quad 
S(k) = -124 + 456k - 519k^2 + 241k^3, \\
\fl G_1(k) = 29952 - 260400k + 932160k^2 - 1727456k^3 
+ 1706012k^4 \\
\lo 
- 810861k^5 + 118529k^6 + 16870k^7, \\
\fl G_2(k) = - \, 3170144 + 37193808k - 190561760k^2 
+ 553816792k^3 \\
\lo 
- 998755638k^4 + 1151765545k^5 - 843129587k^6 \\
\lo
+ 371783811k^7 - 86588015k^8 + 7446900k^9, \\
\fl G_3(k) = - \, 1361952 + 20130192k - 133944304k^2 
+ 527295832k^3 - 1360307178k^4  \\ 
\lo 
+ 2408313485k^5 - 2978361002k^6 + 2565200757k^7 
- 1500811081k^8 \\
\lo 
+ 563283562k^9 - 120201105k^{10} + 10736550k^{11}, \\
\fl G_4(k) = - \, 54723072 + 687806528k - 3817158400k^2 + 
12223132176k^3 \\
\lo 
- 24787677856k^4 + 32872864764k^5 - 28351948208k^6 \\
\lo 
+ 15180036291k^7 - 4446138980k^8 + 454376975k^9 
+ 40055750k^{10}, \\
\fl G_5(k) = 68988096 - 912699392k + 5346987920k^2 
- 18143124368k^3 \\
\lo 
+ 39260441572k^4 - 56288554120k^5 + 53780522743k^6  \\
\lo 
- 33457433369k^7 + 12720106145k^8- 2562547775k^9 \\
\lo 
+ 186064500k^{10}, \\
\fl G_6(k) = 56393856 - 913712256k + 6725558432k^2 
- 29645084416k^3 \\
\lo 
+ 86929567784k^4 - 178232080840k^5 + 261364852406k^6  \\
\lo 
- 275416174516k^7 + 206176050353k^8- 106294583086k^9 \\
\lo 
+ 35472298955k^{10} - 6761687100k^{11} + 538285500k^{12}.
\end{eqnarray*}
The quantity called the resultant determines whether given 
two polynomials have a common root or not. 
Let us consider the two polynomials
\begin{eqnarray}
f(x) = a_0 x^n + a_1 x^{n-1} + \cdots + a_n \quad (a_0 \neq 0), \\
g(x) = b_0 x^m + b_1 x^{m-1} + \cdots + b_m \quad (b_0 \neq 0).
\end{eqnarray}
The resultant of \(f(x)\) and \(g(x)\) is defined by the following 
determinant of order \((m+n)\).
\begin{equation}
R(f,g) = \left|
\begin{array}{ccccccc}
a_0 & a_1 & \cdots & a_n     &             &            & \\
       & a_0 & a_1     & \cdots & a_n      &            & \\
       &        & \cdots & \cdots & \cdots &            & \\ 
       &        &            & a_0     & a_1      & \cdots & a_n \\
b_0 & b_1 & \cdots & b_m     &            &             & \\
       & b_0 & b_1     & \cdots & b_m      &            & \\
       &        & \cdots & \cdots & \cdots &            & \\
       &        & b_0     & b_1      & \cdots & \cdots & b_m 
\end{array}
\right|
\end{equation}
Here, all the blanks represent zeros. 
It is known that the following theorem holds.
\begin{thm}[see, for example, \cite{waerden1991}]\label{thm:resultant}
The two polynomials \(f(x)\) and \(g(x)\) have a common root 
if and only if \(R(f,g) = 0\).
\end{thm}
See \ref{sec:resultant} for the proof. 
The resultant of the two algebraic equations (\ref{eqn:eqn1}) and 
(\ref{eqn:eqn2}) is computed to be~\cite{wolfram1996}
\begin{eqnarray}
\fl - \frac{289340}{4096} \nonumber \\
\fl \times \frac{(k + 2)^9 (k-3)^4(k-4)^2 (2k - 3)^4 (3k - 4)^4
(3k - 1)^4 (5k - 6)^2 (5k - 2)^{29} (7k - 6)^4}
{k^2 (k - 1)^9 (k - 2)^{11} (2k - 1)^{11} (3k - 2)^{16} R(k)^6 S(k)^{10}} 
\nonumber \\ 
\fl \times (29952 - 260400k + 932160k^2 - 1727456k^3 
\nonumber \\
\lo 
+ 1706012k^4 - 810861k^5 + 118529k^6 + 16870k^7) \nonumber \\
\fl \times (- 2787216 + 23021920k - 82567904k^2 
+ 167598204k^3 \nonumber \\
\lo 
- 211881739k^4 + 173025983k^5 - 91304549k^6 \nonumber \\
\lo 
+ 30037885k^7 - 5594600k^8 + 450000k^9)^2, 
\label{eqn:result}
\end{eqnarray}
which does not vanish for \(k \geq 5\). 
Then it is concluded from theorem~\ref{thm:resultant} 
that the two algebraic equations (\ref{eqn:eqn1}) and (\ref{eqn:eqn2}) 
do not have any common solutions. 
Therefore, there exist no common solutions among the \(2k-5\) 
algebraic equations. 
This means that we have no solutions for 
\(\alpha_2=(3k(2k-1)/2)\alpha_0\) (\(\alpha_0=1\)).

\bigskip
\noindent
From the above arguments, we see that the case 5 considered here 
only yields the separable potential of the form
\begin{equation}
V_k = A x^k + B y^k,
\end{equation}
i.e. there exist no new integrable potentials.

\section{A list of integrable homogeneous polynomial potentials}
\label{sec:list}
\noindent
Although the computations in the previous section were performed 
under the implicit assumption that the degree of the potential, 
\(k\), is greater than 4, 
we can carry out the computations for the cases \(k=3\) and \(k=4\) 
in the same manner. 
For \(k=3\), the expressions of \(M_{ij}\) are given by 
\begin{eqnarray*}
\fl M_{11} = \frac{(3\alpha_0 - \alpha_2)
(45\alpha_0 - 2\alpha_2)}{3}, \\
\fl M_{21} = - 30 \alpha_3 (3\alpha_0 - \alpha_2), \quad
M_{22} = - \frac{112\alpha_2 (3\alpha_0 - \alpha_2)}{3} , \\
\fl M_{31} = - 9\alpha_0 \alpha_2 - 7\alpha_2^2 + 45\alpha_3^2, 
\quad
M_{32} = 60 \alpha_2 \alpha_3, \quad
M_{33} = -9(\alpha_0 - 2\alpha_2) (5\alpha_0 - \alpha_2), \\
\fl M_{41} = - 30 \alpha_2 \alpha_3, \quad
M_{42} = - 40 \alpha_2^2, \quad
M_{43} = 30 \alpha_3 (3\alpha_0 - 2\alpha_2), \\
\fl M_{44} = - \frac{8\alpha_2 (3\alpha_0 - 16\alpha_2)}{3}. 
\end{eqnarray*}
Then the condition that the product \(M_{11}M_{22}M_{33}M_{44}\) 
vanishes gives the following relations between \(\alpha_0\) and 
\(\alpha_2\).
\begin{equation}\fl
\alpha_2 = 0, \quad
\alpha_2 = 3\alpha_0, \quad
\alpha_2 = \frac{45}{2} \alpha_0, \quad
\alpha_2 = \frac{1}{2} \alpha_0, \quad
\alpha_2 = 5\alpha_0, \quad
\alpha_2 = \frac{3}{16}\alpha_0.
\end{equation}
When \(\alpha_2=0\), we obtain the separable potential
\begin{equation}\label{eqn:v3a}
V_3 = \alpha_0 x^3 + \alpha_3 y^3
\end{equation}
and when \(\alpha_2 = 3 \alpha_0\) \((\alpha_0=1)\), 
we obtain the potential
\begin{equation}\label{eqn:v3b}
V_3 = x^3 + 3 x y^2 + \alpha_3 y^3,
\end{equation}
which is transformed into the form of (\ref{eqn:v3a}) by the 
rotation of coordinates defined by (\ref{eqn:rotation}). 
When \(\alpha_2=(45/2)\alpha_0\) \((\alpha_0=1)\), 
we obtain an algebraic equation for \(\alpha_3\), which is 
given by (\ref{eqn:eqn1}) with \(k=3\), after the same computations 
as in the previous section. 
Then we obtain the potentials
\begin{eqnarray}
V_3 = x^3 + \frac{45}{2} x y^2 + \frac{17\sqrt{14}\,i}{2} y^3, 
\label{eqn:v3c}\\
V_3 = x^3 + \frac{45}{2} x y^2 - \frac{27\sqrt{3}\,i}{2} y^3. 
\label{eqn:v3d}
\end{eqnarray}
When \(\alpha_2=(1/2)\alpha_0\) \((\alpha_0=1)\), 
we obtain the potential
\begin{equation}\label{eqn:v3e}
V_3 = x^3 + \frac12 x y^2 + \frac{\sqrt{3}\,i}{18} y^3.
\end{equation}
When \(\alpha_2=5\alpha_0\) \((\alpha_0=1)\), 
we obtain the potential
\begin{equation}\label{eqn:v3f}
V_3 = x^3 + 5 x y^2 + \frac{22\sqrt{3}\,i}{9} y^3.
\end{equation}
The potentials (\ref{eqn:v3d}), (\ref{eqn:v3e}), (\ref{eqn:v3f}) 
are transformed into each other by proper rotations of coordinates.
When \(\alpha_2=(3/16)\alpha_0\) \((\alpha_0=1)\), 
we obtain the potential
\begin{equation}\label{eqn:v3g}
V_3 = x^3 + \frac{3}{16} x y^2.
\end{equation}
The potentials (\ref{eqn:v3c}), (\ref{eqn:v3g}) are transformed 
into each other by proper rotations of coordinates.

For \(k=4\), the expressions of \(M_{ij}\) are given by
\begin{eqnarray*}
\fl M_{11} = \frac{(6\alpha_0 - \alpha_2)
(42\alpha_0 - \alpha_2)}{3}, \\
\fl M_{21} = - 11 \alpha_3 (6\alpha_0 - \alpha_2), \quad
M_{22} = - \frac{80\alpha_2 (6\alpha_0 - \alpha_2)}{3}, \\
\fl M_{31} = \frac{540\alpha_0 \alpha_2 - 16\alpha_2^2 + 
99\alpha_3^2 - 1512\alpha_0 \alpha_4 
+ 228\alpha_2 \alpha_4}{6}, \\
\fl M_{32} = - 80 \alpha_3 (3\alpha_0 - \alpha_2), \quad
M_{33} = 80 \alpha_0 \alpha_2, \\
\fl M_{41} = - 2 \alpha_3 (6\alpha_0 + 37\alpha_2 - 48\alpha_4), 
\quad
M_{42} = - \frac{8(32\alpha_2^2 - 27\alpha_3^2 -
 24\alpha_2 \alpha_4)}{3} , \\
\fl M_{43} = 24 \alpha_3 (7\alpha_0 - \alpha_2), \quad
M_{44} = - \frac{64\alpha_2 (3\alpha_0 - 4\alpha_2)}{3}. 
\end{eqnarray*}
Then the condition that the product \(M_{11}M_{22}M_{33}M_{44}\) 
vanishes gives the following relations between \(\alpha_0\) and 
\(\alpha_2\).
\begin{equation}
\alpha_0 = 0, \quad
\alpha_2 = 0, \quad
\alpha_2 = 6\alpha_0, \quad
\alpha_2 = 42\alpha_0, \quad
\alpha_2 = \frac{3}{4}\alpha_0.
\end{equation}
When \(\alpha_0\), we obtain the potential
\begin{equation}\label{eqn:v4a}
V_4 = y^4,
\end{equation}
and when \(\alpha_2=0\) \((\alpha_0=1)\), we obtain the potential
\begin{equation}\label{eqn:v4b}
V_4 = x^4 + \alpha_4 y^4.
\end{equation}
When \(\alpha_2=6\alpha_0\) \((\alpha_0=1)\), we obtain the 
potentials
\begin{eqnarray}
V_4 = x^4 + 6 x^2 y^2 + \alpha_3 x y^3 + \frac{16 + \alpha_3^2}{16} y^4, 
\label{eqn:v4c} \\
V_4 = x^4 + 6 x^2 y^2 + 8 y^4. 
\label{eqn:v4d}
\end{eqnarray}
The potential (\ref{eqn:v4c}) is transformed into the form of 
(\ref{eqn:v4b}) by the rotation of coordinates (\ref{eqn:rotation}).
When \(\alpha_2=42\alpha_0\) \((\alpha_0=1)\), we obtain three 
algebraic equations for \(\alpha_3\), two of which are given by 
(\ref{eqn:eqn1}) and (\ref{eqn:eqn2}) with \(k=4\), after the same 
computations as in the previous section. 
They have common solutions, which is indicated by the fact that 
the resultant (\ref{eqn:result}) vanishes when \(k=4\). 
Then we obtain the potential
\begin{equation}\label{eqn:v4e}
V_4 = x^4 + 42 x^2 y^2 + 28 \sqrt{10}\,i x y^3 - 48 y^4.
\end{equation}
When \(\alpha_2=(3/4)\alpha_0\) \((\alpha_0=1)\), 
we obtain the potential
\begin{equation}\label{eqn:v4f}
V_4 = x^4 + \frac34 x^2 y^2 + \frac18 y^4.
\end{equation}
The potentials (\ref{eqn:v4d}), (\ref{eqn:v4e}), (\ref{eqn:v4f}) are 
transformed into one another by proper rotations of coordinates.

As a consequence, we obtain the following list of integrable 
2D homogeneous polynomial potentials with a polynomial first 
integral of order at most 4  in the momenta. 
\begin{itemize}
\item with a polynomial first integral linear in the momenta
\begin{equation}
V_k = (x^2 + y^2)^{k/2}, \quad k = \mbox{even}.
\end{equation}
\item with a polynomial first integral quadratic in the momenta
\begin{equation}\fl
V_k = \frac{1}{r} \left[ \left (\frac{r+x}{2} \right)^{k+1} 
+ (-1)^k \left (\frac{r-x}{2} \right)^{k+1} \right], \quad 
V_k = A x^k + B y^k.
\end{equation}
\item with a polynomial first integral quartic in the momenta
\begin{equation}\fl
V_3 = x^3 + \frac{3}{16} x y^2, \quad 
V_3 = x^3 + \frac12 x y^2 + \frac{\sqrt{3}\,i}{18} y^3, \quad
V_4 = x^4 + \frac{3}{4} x^2 y^2 + \frac{1}{8} y^4.
\end{equation}
\end{itemize}
As far as the authors know, no one has discovered any polynomial 
first integral which is {\em genuinely} quintic or higher orders 
in the momenta. 
It is still an open problem whether or not there exist such 
polynomial first integrals. 

\ack
The authors are grateful to J.~Hietarinta, A.~Ramani and  
B.~Grammaticos for their valuable comments. 
We also thank the anonymous referee for providing 
helpful suggestions to the original version of the introduction. 
This work was partially supported by the Grants-in-Aid for Scientific 
Research of Japan Society for the Promotion of Science (JSPS), 
No.~11304004 and No.~12640227.

\appendix
\section{Proof of the properties 1 and 2}\label{sec:method}
\subsection{Time reflection symmetry -- proof of the property 1}
\noindent
The system (\ref{eqn:system}) is invariant by the time reflection
\begin{equation}
t \to -t, \quad
x \to x, \quad y \to y, \quad
p_x \to - p_x, \quad p_y \to - p_y.
\end{equation} 
If \({\mathit \Phi}(x,y,p_x,p_y)\) is a first integral of the system 
(\ref{eqn:system}), then \({\mathit \Phi}(x,y,- p_x,- p_y)\) is 
also a first integral because of the time reflection symmetry of 
the system. 
Note here that every first integral \({\mathit \Phi}\) can be 
decomposed into the sum of \({\mathit \Phi}_{\rm even}\) and 
\({\mathit \Phi}_{\rm odd}\), given by 
\begin{equation}
{\mathit \Phi}_{\rm even} = \frac{{\mathit \Phi}(x,y,p_x,p_y) 
+ {\mathit \Phi}(x,y,-p_x, - p_y)}{2}
\end{equation}
and
\begin{equation}
{\mathit \Phi}_{\rm odd} = \frac{{\mathit \Phi}(x,y,p_x,p_y) 
- {\mathit \Phi}(x,y,-p_x, - p_y)}{2}.
\end{equation}
We can see that \({\mathit \Phi}_{\rm even}\) is a first integral 
which is even in the momenta and that \({\mathit \Phi}_{\rm odd}\) 
is a first integral which is odd in the momenta. 
That is, if \({\mathit \Phi} 
={\mathit \Phi}_{\rm even}+{\mathit \Phi}_{\rm odd}\) 
is a first integral of the system (\ref{eqn:system}), 
then \({\mathit \Phi}_{\rm even}\) and \({\mathit \Phi}_{\rm odd}\) 
are also first integrals. 
Therefore, we can assume that a first integral is either 
even or odd in the momenta from the beginning.

\subsection{Scale invariance -- proof of the property 2}
\noindent
The system (\ref{eqn:system}) with a homogeneous polynomial 
potential of degree \(k\) is invariant by the scale transformation 
\begin{equation}\label{eqn:hscale}
\fl
t \to \sigma^{-1} t, \quad 
x \to \sigma^{2/(k-2)} x,\quad y \to \sigma^{2/(k-2)} y, \quad
p_x \to \sigma^{k/(k-2)} p_x, \quad p_y \to \sigma^{k/(k-2)} p_y. 
\end{equation}
In general, a system of differential equations
\begin{equation}\label{eqn:diffeqs}
\frac{dx_i}{dt} = F_i(x_1,x_2,\ldots,x_n),\quad (i = 1,2,\ldots,N)
\end{equation}
is called a scale invariant system if it is invariant by 
the scale transformation
\begin{equation}\label{eqn:scale}
t \to \sigma^{-1} t, \quad x_i \to \sigma^{g_i} x_i, 
\quad (i =1,2,\ldots,N)
\end{equation}
with an arbitrary parameter \(\sigma\) and proper constants \(g_i\). 
A function \({\mathit \Phi}(x_1,x_2,\ldots,x_N)\) is said to be  
weighted-homogeneous with a weight \(M\) if it satisfies
\begin{equation}
{\mathit \Phi
}(\sigma^{g_1}x_1,\sigma^{g_2}x_2,\ldots,\sigma^{g_N}x_N) 
= \sigma^M {\mathit \Phi}(x_1,x_2,\ldots,x_N).
\end{equation}
Suppose now that the scale invariant system (\ref{eqn:diffeqs}) 
has a polynomial first integral \({\mathit \Phi}\), 
which can be written in the form
\begin{equation}\label{eqn:sumPhi}
{\mathit \Phi} = \sum_m {\mathit \Phi}_m,
\end{equation}
where each polynomial \({\mathit \Phi}_m\) is weighted-homogeneous 
with a weight \(m\). The scale transformation (\ref{eqn:scale}) 
transforms the first integral (\ref{eqn:sumPhi}) into 
\begin{equation}\label{eqn:sumPhi'}
{\mathit \Phi}' = \sum_m \sigma^m {\mathit \Phi}_m,
\end{equation}
which is again a first integral for an arbitrary \(\sigma\) 
because of the scale invariance of the system. 
Then, it is concluded that each polynomial \({\mathit \Phi}_m\) 
is a first integral. 
Therefore, we can assume that a polynomial first integral is 
weighted-homogeneous from the beginning.

\section{Resultant -- proof of theorem \ref{thm:resultant}}
\label{sec:resultant}
\noindent
The resultant is an algebraic tool for elimination of a variable 
between two algebraic equations and gives the condition that 
the two algebraic equations have a common root 
(theorem~\ref{thm:resultant}). 
See \cite{waerden1991} for more details of the resultant and 
its applications. %Here is a proof of theorem~\ref{thm:resultant}. 

\bigskip
\noindent
{\bf Proof of theorem~\ref{thm:resultant}:}
Suppose that the two polynomials have a common root, say \(\alpha\). 
Then the following simultaneous algebraic equations hold.
\begin{equation}\label{eqn:simforR}
\left\{
\begin{array}{rl}
\alpha^{m-1} f(\alpha) &= a_0 \alpha^{m+n-1} + a_1 \alpha^{m+n-2} 
+ \cdots + a_n \alpha^{m-1} = 0 \\
\alpha^{m-2} f(\alpha) &= a_0 \alpha^{m+n-2} + a_1 \alpha^{m+n-3} 
+ \cdots + a_n \alpha^{m-2} = 0 \\
& \,\, \vdots \\
f(\alpha) &= a_0 \alpha^{n} + a_1 \alpha^{n-1} + \cdots + a_n = 0 \\
\alpha^{n-1} g(\alpha) &= b_0 \alpha^{m+n-1} + b_1 \alpha^{m+n-2} 
+ \cdots + b_m \alpha^{n-1} = 0 \\
\alpha^{n-2} g(\alpha) &= b_0 \alpha^{m+n-2} + b_1 \alpha^{m+n-3} 
+ \cdots + b_m \alpha^{n-2} = 0 \\
& \,\, \vdots \\
g(\alpha) &= b_0 \alpha^{m} + b_1 \alpha^{m-1} + \cdots + b_m = 0
\end{array}
\right.
\end{equation}
If we multiply the \(l\)th column of the resultant by \(\alpha^l\) 
and add them to the \((m+n)\)-th column, then all the elements of 
the \((m+n)\)-th column of the resultant vanish because of 
(\ref{eqn:simforR}). Therefore, \(R(f,g)=0\).

Let us next assume that \(R(f,g)=0\). Then the \(m+n\) row 
vectors of the resultant are linear dependent. Let
\(\mbox{\boldmath $a$}_1, \mbox{\boldmath $a$}_2, \ldots,
\mbox{\boldmath $a$}_m, \mbox{\boldmath $b$}_1, 
\mbox{\boldmath $b$}_2, \ldots, \mbox{\boldmath $b$}_n\) 
be the row vectors of the resultant. Then the relation
\begin{equation}\label{eqn:lindep}
\sum_{i=1}^m c_i \mbox{\boldmath $a$}_i +
\sum_{j=1}^n d_j \mbox{\boldmath $b$}_j = 0
\end{equation}
holds with \((c_1,\ldots,d_n) \neq (0,\ldots,0)\). 
Multiplying the \(l\)th element of (\ref{eqn:lindep}) by \(x^{m+n-l}\) 
and adding them up, we have
\begin{equation}
\sum_{i=1}^m c_i x^{m-i} f(x) + \sum_{j=1}^n d_j x^{n-j} g(x) = 0,
\end{equation}
which is reduced to
\begin{equation}\label{eqn:degeq}
h(x) f(x) = - k(x) g(x),
\end{equation}
where 
\(\sum_{i=1}^m c_i x^{m-i} = h(x),\sum_{j=1}^n d_j x^{n-j} = k(x)\). 
Let \(\deg f(x)\) denote the degree of a polynomial \(f(x)\). 
Then, we have the relation
\begin{equation}\label{eqn:degineq}
\deg k(x) \leq n-1 < n = \deg f(x).
\end{equation} 
If \(f(x)\) and \(g(x)\) have no common factors, then \(f(x)\) must be 
a factor of \(k(x)\) because of (\ref{eqn:degeq}). 
This contradicts equation~(\ref{eqn:degineq}). 
Therefore, \(f(x)\) and \(g(x)\) have at least one common factor, i.e. 
they have at least one common root. 
This completes the proof of theorem~\ref{thm:resultant}.

\end{document}